# FOURIER DIFFUSION AND SPECIAL RELATIVITY IN NANOTECHNOLOGY

Janina Marciak-Kozlowska

Institute of Electron Technology, Al. Lotnikow 32/46 ,02-668 Warsaw Poland

and

Miroslaw Kozlowski

Abstract

In this paper the transport phenomena in On-Chip-Transmission Line are investigated . The transport equation are developed and solved. The near light speed phenomena in OCTL are investigated

Introduction

The nanotechnology industry has evolved at very high rate , particularly in the past ten years. transistor channel length have decreased from 2.0 um in 1980 to 0.5 um in 1992 to current (2006)systems with channel length ~ 50nm. Recently the International Technology Road Map for Semiconductors (ITRS) was postulated. The IRST is devoted to the study of the proposed of the global interconnects and problems with the time delays of the transistors and logical systems. One of the *remedium* for system time delays is the technology On –Chip Transmission Lines. The On-Chip transmission Technology allows for the electron transmission with near- of – light On- Chip electrical interconnection
This paper is addressed to the thermal transport in On-Chip –Transmission Lines.

In the first paragraph we develop the theoretical framework for heat transport in On-Chip- Transmission Line (OCTL) We formulate the transmission line heat transport equation and solved it for the Cauchy boundary conditions.

In the second paragraph we study the special relativity influence on the information transmission in OCTL. We show that the standard Fourier approximation leads to infinite speed of the data transmission. On the other hand the hyperbolic diffusion equation formulated in paragraph 1 gives the finite speed of the transmission of the data

1. Thermal diffusion in OCTL

Dynamics of nonequilibrium electrons and phonons in metals, semiconductors have been the focus of much attention because of their fundamental interest in solid state physics and nanotechnology.

In metals, relaxation dynamics of optically excited nonequilibrium electrons has been extensively studied by pump – probe techniques using femtosecond lasers [1 – 4].

Recently [5] it was shown that the optically excited metals relax to equilibrium with two models: rapid electron relaxation and slow thermal relaxation through the creation of the optical phonons. The same processes will occur in OCTL

In this paper we develop the hyperbolic thermal diffusion equation with two models: electrons and phonons relaxation. These two modes are characterized by two relaxation times $\tau_1$ for electrons and $\tau_2$ for phonons. This new equation is the generalization of our one mode hyperbolic equation, with only electrons degrees of freedom, $\tau_1$ [ 6 ]. The hyperbolic two mode equation is the analogous equation to Klein – Gordon equation and allows the heat propagation with finite speed.

As was shown in paper [6] for high frequency laser pulses the diffusion velocity exceeds that of light. This is not possible and merely demonstrates that Fourier equation is not really correct. Oliver Heaviside was well aware of this writing [7] :

*All diffusion formulae (as in heat conduction) show instantaneous action to an infinite distances of a source, though only to an infinitesimal extent. To make the theory of heat diffusion be rational as well as practical some modification of the equation to remove the instantaneity, however little difference it may make quantatively, in general.*

August 1876 saw the appearance in Philosophical Magazine [ 7 ] the paper which extended the mathematical understanding the diffusion (Fourier) equation. O. Heaviside for the first time wrote the hyperbolic diffusion equation for the voltage $V(x,t)$. assuming a uniform resistance, capacitance and inductance per unit length, $k$, $c$ and $s$ respectively he arrived at:

$$\frac{\partial^2 V(x,t)}{\partial x^2} = kc\frac{\partial V(x,t)}{\partial t} + sc\frac{\partial^2 V(x,t)}{\partial t^2} \qquad (1.1)$$

The discussion of the broad sense of the Heaviside equation (1) can be find out, for example in our monograph [ 8], viz,

$$\tau^2 \frac{\partial^2 T}{\partial t^2} + \tau \frac{\partial T}{\partial t} + \frac{2V\tau}{\hbar}T = \tau \frac{\hbar}{m}\nabla^2 T \qquad (1.2)$$

Equation (1.2) is the heat transport equation for the transmission line .In Eq. (1.2) $T(\vec{r},t)$ denotes the temperature field, $V$ is the external potential, $m$ is the mass of heat carrier and $\tau$ is the relaxation time

$$\tau = \frac{\hbar}{mv^2} \qquad (1.3)$$

As can be seen from formulae (1.2) and (1.3) in hyperbolic diffusion equation the same relaxation time $\tau$ is assumed for both type of motion: wave and diffusion.

This can not be so obvious. For example let us consider the simpler harmonic oscillator equation:

$$m\frac{d^2 x}{dt^2} + kx + c\frac{dx}{dt} = 0 \qquad (1.4)$$

Equation (4) can be written as

$$\tau^2 \frac{d^2 x}{dt^2} + \tau \frac{dx}{dt} + x = 0 \qquad (1.5)$$

where

$$\tau^2 = \frac{m}{k}, \qquad \tau = \frac{c}{k} \qquad (1.6)$$

i.e.

$$c^2 = km \qquad (1.7)$$

As it was well known equation (1.5) with formula (1.6) describes only the weakly damped (periodic) motion of the harmonic oscillator (HO). It must be stressed that for HO exists also critically damped and overdamped modes which are not describes by the equation (1.5). The general master equation for HO must be written as

$$\tau_1^2 \frac{d^2 x}{dt^2} + \tau_2 \frac{dx}{dt} + x = 0 \tag{1.8}$$

Following the discussion of the formulae (1.5) to (1.8) we argue that the general hyperbolic diffusion equation can be written as:

$$\tau_1^2 \frac{\partial^2 T}{\partial t^2} + \tau_2 \frac{\partial T}{\partial t} + \frac{2V\tau_2}{\hbar} T = \frac{\hbar}{m} \tau_2 \nabla^2 T \tag{1.9}$$

and $\tau_1 \neq \tau_2$

Equation (9) describes the temperature field generated by ultra-short laser pulses. In Eq. (1.9) two modes: wave and diffusion are described by different relaxation times.

For quantum hyperbolic equation (1.9) we seek solution in the form (in 1D)

$$T(x,t) = e^{-\frac{t\tau_2}{2\tau_1^2}} u(x,t) \tag{1.10}$$

After substitution Eq. (1.10) into Eq. (1.9) one obtains

$$\frac{1}{v^2} \frac{\partial^2 u}{\partial t^2} - \frac{\partial^2 u}{\partial x^2} + qu(x,t) = 0 \tag{1.11}$$

where

$$v^2 = \frac{\hbar \tau_2}{m \tau_1^2}, \qquad q = \left( \frac{2Vm}{\hbar^2} - \frac{1}{4} \frac{m}{\hbar} \frac{\tau_2}{\tau_1^2} \right) \tag{1.12}$$

Equation (1.11) is the thermal two-mode Klein – Gordon equation and is the generalization of Klein – Gordon one mode equation developed in our monograph.

For Cauchy initial condition

$$u(x,0) = f(x), \qquad \frac{\partial u(x,0)}{\partial t} = g(x) \tag{1.13}$$

the solution of Eq. (1.11) has the form

$$u(x,t) = \frac{f(x-vt) + f(x+vt)}{2}$$
$$+ \frac{1}{2v} \int_{x-vt}^{x+vt} g(\varsigma) I_0 \left[ \sqrt{-q(v^2 t^2 - (x-\varsigma)^2)} \right] d\varsigma$$
$$+ \frac{(v\sqrt{-q})t}{2} \int_{x-vt}^{x+vt} f(\varsigma) \frac{I_1 \left[ \sqrt{-q(v^2 t^2 - (x-\varsigma)^2)} \right]}{\sqrt{v^2 t^2 - (x-\varsigma)^2}} d\varsigma \tag{1.14}$$

for $q < 0$

and

$$u(x,t) = \frac{f(x-vt)+f(x+vt)}{2}$$

$$+ \frac{1}{2v}\int_{x-vt}^{x+vt} g(\varsigma) J_0\left[\sqrt{q(v^2t^2-(x-\varsigma)^2)}\right] d\varsigma$$

$$- \frac{(v\sqrt{q})t}{2}\int_{x-vt}^{x+vt} f(\varsigma) \frac{J_1\left[\sqrt{q(v^2t^2-(x-\varsigma)^2)}\right]}{\sqrt{v^2t^2-(x-\varsigma)^2}} d\varsigma$$

(1.15)

for $q > 0$.

3. Minkowski space time and diffusion

In this paragraph we develop the description of the heat transport in Minkowski spacetime of the OCTS. In the context special relativity theory we investigate the Fourier equation and hyperbolic diffusion equation. We calculate the speeds of the heat diffusion in Fourier approximation and show that for high energy laser beam the heat diffusion exceeds the light velocity. We show that this results breaks the causality of the thermal phenomena in Minkowski spacetime. The same phenomena we describe in the framework of hyperbolic heat diffusion equation and show that in that case speed of diffusion is always smaller than light velocity.

We may use the concept that the speed of light *in vacuo* provides an upper limit on the speed with which a signal can travel between two events to establish whether or not any two events could be connected. In the interest of simplicity we shall work with one space dimension $x_1 = x$ and the time dimension $x_o = ct$ of the Minkowski spacetime,. Now let us consider events (1) and (2): their Minkowski interval $\Delta s$ satisfies the relationship:

$$\Delta s^2 = c^2 \Delta t^2 - \Delta x^2 \qquad (2.1)$$

Without loss of generality we take Event 1 to be at $x = 0$, $t = 0$. Then Event 2 can be only related to Event 1 if it is possible for a signal traveling at the speed of light, to connect them. Let the, Event 2 is at $(\Delta x, c\Delta t)$, its relationship to Event 1 depending on whether $\Delta s > 0, = 0,$ or $< 0$.

We may summarize the three possibilities as follows:

Case A  **timelike** *interval*, $|\Delta x_A| < c\Delta t$, or $\Delta s^2 > 0$. Event 2 can be related to Event 1, Events 1 and 2 can be in causal relation.

Case B   **lightlike** interval, $|\Delta x_B| = c\Delta t$, or $\Delta s^2 = 0$. Event 2 can only be related to Event 1 by a light signal.

Case C   **spacelike** interval $|\Delta x_A| > c\Delta t$, or $\Delta s^2 < 0$. Event 2 cannot be related to Event 1, for in that case $v > c$.

Now let us consider the case C in more details. At first sight it seems that in case C we can find out the reference frame in which two Events $c^>$ and $c^<$ always fulfils the relations $t_{c^>} - t_{c^<} > 0$. but it is not true. For let us choose the inertial frame U' in which $t_{c^>} - t_{c^<} > 0$. In reference frame U which is moving with speed $V$ relative to U', where

$$V = c \frac{c(t'_{c^>} - t'_{c^<})}{x'_{c^<} - x'_{c^>}} \qquad (2.2)$$

Speed $V<c$ for

$$\left| \frac{c(t'_{c^>} - t'_{c^<})}{x'_{c^<} - x'_{c^>}} \right| < 1 \qquad (2.3)$$

Let us calculate $t_{c^>} - t_{c^<}$ in the reference frame U

$$t_{c^>} - t_{c^<} = \frac{1}{\sqrt{1 - \frac{V^2}{c^2}}} \left[ \frac{V}{c^2}(x'_{c^>} - x'_{c^<}) + (t'_{c^>} - t'_{c^<}) \right] =$$

$$\frac{1}{\sqrt{1 - \frac{V^2}{c^2}}} \left[ \frac{t'_{c^>} - t'_{c^<}}{x'_{c^>} - x'_{c^<}}(x'_{c^>} - x'_{c^<}) + (t'_{c^>} - t'_{c^<}) \right] = 0 \qquad (2.4)$$

For the greater $V$ we will have $t_{c^>} - t_{c^<} < 0$. It means that for the spacelike intervals the sign of $t_{c^>} - t_{c^<}$ depends on the speed $V$, i.e. causality relation for spacelike events is not valid.

2.1  Fourier diffusion and special relativity

In paper [9] the speed of the diffusion signals was calculated

$$v = \sqrt{2D\omega} \qquad (2.5)$$

where

$$D = \frac{\hbar}{m} \qquad (2.6)$$

and $\omega$ is the angular frequency of the laser pulses. Considering formula (5) and (6) one obtains

$$v = c\sqrt{2\frac{\hbar\omega}{mc^2}} \quad (2.7)$$

and $v \geq c$ for $\hbar\omega \geq mc^2$.

From formula (7) we conclude that for $\hbar\omega > mc^2$ the Fourier diffusion equation is in contradiction with special relativity theory and breaks the causality in transport phenomena.

2.2 Hyperbolic diffusion and special relativity

In monograph [10] the hyperbolic model of the heat transport phenomena was formulated. It was shown that the description of the ultrashort thermal energy transport needs the hyperbolic diffusion equation (one dimension transport)

$$\tau\frac{\partial^2 T}{\partial t^2} + \frac{\partial T}{\partial t} = D\frac{\partial^2 T}{\partial x^2} \quad (2.9)$$

In the equation (9) $\tau = \dfrac{\hbar}{m\alpha^2 c^2}$ is the relaxation time, $m$ = mass of the heat carrier, $\alpha$ is the coupling constant and $c$ is the light speed in vacuum, $T(x,t)$ is the temperature field and $D = \hbar/m$.

In paper [1] the speed of the thermal propagation $v$ was calculated

$$v = \frac{2\hbar}{m}\sqrt{-\frac{m}{2\hbar}\tau\omega^2 + \frac{m\omega}{2\hbar}(1+\tau^2\omega^2)^{1/2}} \quad (2.10)$$

Considering that $\tau = \hbar/m\alpha^2 c^2$ formula (10) can be written as

$$v = \frac{2\hbar}{m}\sqrt{-\frac{m}{2\hbar}\frac{\hbar\omega^2}{mc^2\alpha^2} + \frac{m\omega}{2\hbar}(1+\frac{\hbar^2\omega^2}{m^2c^4\alpha^4})^{1/2}} \quad (2.11)$$

For

$$\frac{\hbar\omega}{mc^2\alpha^2} < 1, \quad \frac{\hbar\omega^2}{mc^2} < 1 \quad (2.12)$$

one obtains from formula (11)

$$v = \sqrt{\frac{2\hbar}{m}\omega} \quad (2.13)$$

Formally formula (13) is the same as formula (7) but considering inequality (11) we obtain

$$v = \sqrt{\frac{2\hbar\omega}{m}} = \sqrt{2}\alpha c < c \qquad (2.14)$$

and causality is not broken.

For

$$\frac{\hbar\omega}{mc^2} > 1; \qquad \frac{\hbar\omega}{\alpha^2 mc^2} > 1 \qquad (2.15)$$

we obtain from formula (11)

$$v = \alpha c, \qquad v < c \qquad (2.16)$$

Considering formulae (14) and (16) we conclude that the hyperbolic diffusion equation (9) describes the thermal phenomena in accordance with special relativity theory and causality is not broken independently of laser beam energy.

When the amplitude of the laser beam approaches the critical electric field of quantum electrodynamics (Schwinger field [3]) the vacuum becomes polarized and electron – positron pairs are created in vacuum [3]. On a distance equal to the Compton length, $\lambdabar_C = \hbar/m_e c$, the work of critical field on an electron is equal to the electron rest mass energy $m_e c^2$, i.e. $eE_{Sch}\lambdabar_C = m_e c^2$. The dimensionless parameter

$$\frac{E}{E_{Sch}} = \frac{e\hbar E}{m_e^2 c^3} \qquad (2.17)$$

becomes equal to unity for electromagnetic wave intensity of the order of

$$I = \frac{c}{r_e \lambdabar_C^2} \frac{m_e c^2}{4\pi} \cong 4.7 \cdot 10^{29} \tfrac{W}{cm^2} \qquad (2.18)$$

where $r_e$ is the classical electron radius [12]. For such ultra high intensities the effects of nonlinear quantum electrodynamics plays a key role: laser beams excite virtual electron – positron pairs. As a result the vacuum acquires a finite electric and magnetic susceptibility which lead to the scattering of light by light. The cross section for the photon – photon interaction is given by:

$$\sigma_{\gamma\gamma \to \gamma\gamma} = \frac{973}{10125} \frac{\alpha^3}{\pi^2} r_e^2 \left(\frac{\hbar\omega}{m_e c^2}\right)^6, \qquad \text{for}$$

$\hbar\omega / m_e c^2 < 1$ and reaches its maximum, $\sigma_{max} \approx 10^{-20} cm^2$ for $\hbar\omega \approx m_e c^2$ [11].

Considering formulae (18) and (19) we conclude that linear hyperbolic diffusion equation is valid only for the laser intensities $I \leq 10^{29}$ W/cm$^2$. for high intensities the nonlinear hyperbolic diffusion equation must be formulated and solved.

Table 1

Hierarchical structure of the thermal excitation

| Interaction | $\alpha$ | $mc^2\alpha$ |
|---|---|---|
| Electromagnetic | $137^{-1}$ | $0.5/137$ |
| Strong | $\dfrac{15}{100}$ | $\dfrac{140 \cdot 15}{100}$ for pions |
|  |  | $\dfrac{1000 \cdot 15}{100}$ for nucleons |
| Quark - Quark | 1 | $417^*$ |

- D.H. Perkins, Introduction to high energy physics, Addison – Wesley, USA 1987

9/23/2006